# Ultrafast Insight into High energy (C, D) Excitons in Few Layer WS$_2$

Tanmay Goswami[†], Himanshu Bhatt[†], K. Justice Babu[†], Gurpreet Kaur[†], Nandan Ghorai[†], Hirendra N. Ghosh*[,†,ⴕ]

[†]*Institute of Nano Science and Technology, Mohali, Punjab 160062, India*

[ⴕ]*Radiation and Photochemistry Division, Bhabha Atomic Research Centre, Mumbai 400085, India*

**Abstract:** High energy (C, D) excitons possess remarkable influence over the optical properties of layered transition metal dichalcogenides (TMDCs) and comprehensive understanding of these may have revolutionary effect on 2D opto-electronic devices. Herein, we employed transient absorption spectroscopy to monitor the underlying photo-physical processes involved with C, D excitons in few layer WS$_2$. We observed a strong inter-valley coupling across the momentum space. C, D dynamics were significantly slower as compared to canonical A, B excitons, as a consequence of the indirect Λ-Γ relaxation in C, D, unlike K-K direct combination in A, B. Optical behaviour of D excitons was found to be more like A, B, contrary to C, which enjoy unique band nesting effects. Also, C excitons do not hold in any specific position of the momentum space, rather depends upon the photon energy. All these excitons immensely influence each other irrespective of the excitation energy.

## Introduction:

Opto-electronic response of any material has profound reliance on the efficient absorption of electromagnetic radiation and subsequent relaxation of the photogenerated hot carriers in the system[1,2]. The efficacy of an quantum system in various opto-electronic applications like light-emitting diodes[3], photodetectors[4,5], opto-valleytronics[6,7] or energy harvesting devices[8] could possibly be improved by slowing down the relaxation of hot carriers or by exploiting broader absorption range. Detailed understanding of the excitonic features generated in an extended range of the radiation would be crucial for developing advanced photonic devices out of any material.

Nowadays, layered Transition Metal Dichalcogenides (TMDCs) are being projected as semiconducting equivalent of Graphene[9,10] and they confer great potential towards ultrathin and flexible opto-electronic devices[11,12,13]. As the 2D structure of TMDC approaches the monolayer limit, diverse electronic and optical properties emerge in the system, like exceptionally strong coulombic interactions[14], increment in the electronic band gap[15,16], strong spin-orbit coupling[17,18,19], efficient light-matter interactions and fascinating spin-valley physics[20,21]. The optical absorption spectrum of 2D TMDC materials are generally comprised of four excitonic resonances usually labelled as A, B, C, D in energetic order[22,23], however the spectrum is dominated by the C, D transitions, suggested to be formed in a parallel band structure arrangement in their density of states and largely affected by band nesting phenomena[24,25]. The oscillator strengths of these excitons are significantly higher than those low energy ones, even greater than monolayer graphene[26] yet, the quantum efficiency of these excitons was found to be very low[24]. This indicates that the relaxation process associated with these high energy excitons must be very different from those of fundamental excitons (A, B). Dominance of phonon mediated hot carrier relaxation could be one of the key reasons behind this poor quantum efficiency. D. Kozawa et al. reported that the photogenerated carriers (electron/hole) in the band-nesting region exhibit fast intraband relaxation towards nearest excited states ($\Lambda$ valley/$\Gamma$ hill), followed by a relatively slow non-radiative decay to reach K/K′ state in monolayer $MoS_2$[24]. They attributed this fast relaxation to the spontaneous charge separation in the momentum space (Band Nesting effect), which eventually restricts the quantum efficiency of the C excitonic state. Moreover, as 2D systems undergo bulk to monolayer transition, the relaxation processes are modified accordingly, which may be attributed to the increased carrier interactions owing to reduced coulomb screening, bandgap renormalization and indirect to direct band gap transition. Hence, it is very important to study these excitonic species in different dimensions. N. Kumar et al. employed transient absorption microscopy in bulk $MoS_2$, where they observed efficient inter-valley carrier transfer (K to Γ) of K excitons due to weak exciton binding energy in bulk (25 meV) and significant indirect gap effect in the energy band structure[27]. In 2015, T. Borzada et al. reported for the very first time the relaxation dynamics of C excitons along with A, B in case of few layer $MoS_2$[23]. Later on, few other articles surfaced where C-exciton dynamics were studied and quantified in monolayer $MoS_2$. L. Wang et al. observed very slow C exciton relaxation in the parallel energy band arrangement and discussed this behaviour as a consequence of excited state coulombic nature along K and Γ space[28]. Y. Li. et al. reported relatively faster C exciton dynamics in monolayer $MoS_2$ attributing to the slow inter-valley carrier transfer[29]. However, the formation characteristics of C excitons has never been discussed in femtosecond time scale. Also, there are no reports regarding D excitonic behaviour in these layered materials. D exciton does have individual identity in the steady state absorption and is expected to bear significant independent effect over optical behaviours in these materials. Henceforth, understanding the nature of these high energy excited states and underlying photophysical processes in reduced dimensionality could be of key importance towards better implications of these materials in photonic devices.

In this communication we present a spectroscopic overview of the high energy C, D excitons in the mirror of A, B in few layer $WS_2$, elucidating formation as well as relaxation dynamics ranging in a broad UV-Visible region. We observed low intensity high energy C, D excitonic bleach signals along with A, B in the transient absorption spectra. We explored the exciton dynamics of D band, which is very different as compared to that of C exciton and it behaves very similar to the fundamental A, B excitons. Detailed spectral analysis proposed a tentative position for D excitons, closer to Γ and not in a parallel band arrangement like C exciton. We also found that there is no particular momentum space position for C excitons and these are strongly reliant on the excitation photon energy. We elucidated that, in few layer $WS_2$ C, D excitons follow Λ-Γ indirect relaxation pathway, contrary to direct K-K recombination of A, B. Both C, D relaxations were found to retarded in nature due to the involvement of inter-valley scattering process as well as state filling induced Pauli blocking effect in K and Λ valley.

## Results

**Basic characterisation and native photo-physical features in few layer $WS_2$.** $WS_2$ few layers were exfoliated in N-Methyl Pyrrolidone (NMP) using Liquid exfoliation method, a simple, inexpensive, and mass productive technique in comparison to the generally used chemical vapour deposition (CVD) or mechanical exfoliation methods. Raman spectra of few layer $WS_2$ are shown in **supplementary Figure 1**. From AFM study we found that $WS_2$ sheets are comprised of 5-6 no of layers (**Supplementary Figure 2**). The optical signature of a material reveals the basic features of their energy band structure. Figure 1a displays the absorption spectra of few layer $WS_2$, presenting all four excitonic peaks, i.e., A, B, C, D at around 636, 530, 462 and 423 nm respectively (Peaks are labelled following their usual convention[23,30]). Quantum confinement with lower dimensionality and strong coulomb interaction ensures dominance of excitonic features in the steady state absorption spectra. Here, the spectrum is primarily subjugated by strong absorptions from high energy excitons (C, D), which is quite common in low-dimensional TMDC materials.

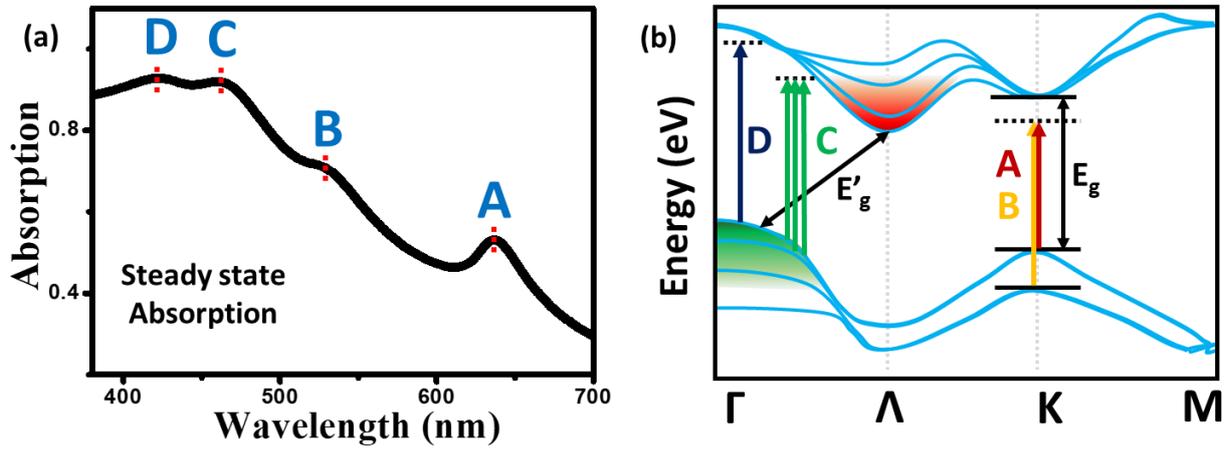

**Figure 1: Steady state optical characteristics of few layer WS$_2$. (a)** Steady state absorption spectra of few layer WS$_2$. **(b)** Quasiparticle optical band structure of few layer WS$_2$, adapted from previous reports on TMDC[25,31]. E$_g$ and E'$_g$ represent direct and indirect optical band gap (In few layer E$_g$ > E'$_g$).

In Figure 1b we discussed a simplified optical band diagram for few layer WS$_2$, representing A, B, C, D excitonic transitions. In layered TMDC systems, interlayer interaction plays significant role in defining electronic band structures of the system. In monolayer, conduction band minima (CBM) and valence band maxima (VBM) both lie in K/K' region of the Brillouin zone. However, with increasing no of layers Λ valley shifts downwards and Γ hill shifts upwards, leaving K/K' valley or hills almost unaltered, as a result of which CBM now resides in Λ valley and VBM in Γ hill[31,32]. A and B peaks originate from the excitonic transitions at the K/K' spaces between conduction band minima and spin splitted valence band maxima[13,33]. On the other hand, the high energy exciton (C) arises from nearly-degenerate excitonic states associated with the van Hove singularity in the joint density of state (JDOS) diagram, between Λ and Γ space[29,34]. The locally parallel bands near Λ space gives rise to local minima in optical band structure with the formation of C excitons[34]. However, we could not find any specific report regarding momentum space position for the D excitonic state. S. H. Aleithan et al. suggested that the D excitonic state could be originating from a higher energy conduction band valley close to C, or could be the result of another band nesting effect in JDOS[35]. They anticipated that, C and D excitonic states would also have bound states shifted by a binding energy ~ 500 meV below conduction band minima in the optical band structure, just like A and B excitons. Here, we employed transient absorption spectroscopy to get a closer look on these high energy excitons alongside A, B in ultrafast time scale.

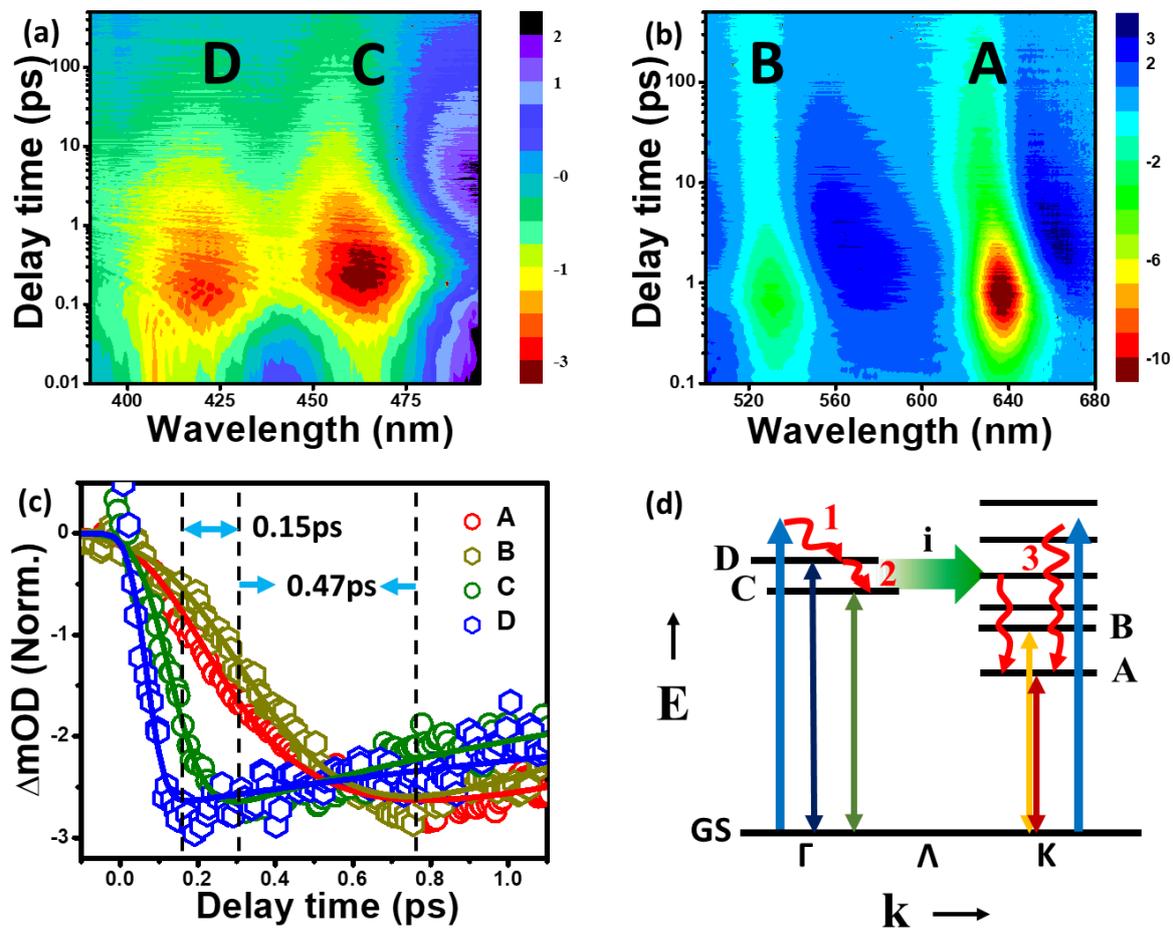

**Figure 2: Excitonic bleach in transient spectroscopy. (a, b)** 2D colour contour plot of D, C, B, A excitonic features representing their spectral and temporal evolution in few layer $WS_2$, with 370 nm (3.3 eV) photoexcitation and 100 μJ/cm$^2$ excitation fluence. **(c)** Normalized growth profiles of all four excitonic features probed at bleach maxima at 636 nm, 532 nm, 464 nm and 422 nm for D, C, B, A exciton respectively. **(d)** Simple energy diagram

representing probable thermalization of hot plasma into corresponding excitonic states. 1, 2, 3 are intra-valley thermalization associated with D, C and A excitons respectively. 'i' denotes inter-valley electron transfer from C, D towards A, B states.

**Evolution of excitonic features.** Transient responses were collected with non-resonant high energy excitation (3.3 eV) probing at different pump-probe delay time and displayed in Figure 2a, b. There are four photo-induced transient bleaches appeared in 2D colour contour plot peaking at around 636, 532, 464 and 422 nm, corresponds to A, B, C and D excitonic transition, as discussed in steady state absorption spectra in Figure 1a. Herein 370 nm pump provides sufficient photon energy to generate different excitonic states of $WS_2$ as shown in Figure 2a and Figure 2b. The intensity of C and D excitonic bleach signal is significantly lower as compared to that of A and B excitonic bleach signals **(See Supplementary Figure 4 for more clear picture),** despite possessing higher oscillator strength[26]. The spectral amplitude (S) of a transient feature is necessarily proportional to the product of injected carrier concentration / exciton density in a particular state (n) and oscillator strength of the corresponding optical transition (f), i.e., S α n.f[36]. Hence, the reason behind this lower spectral weight in the high energy regime could be due to lower carrier density. Gradual enhancement of C exciton intensity with increment of carrier density in monolayer $MoS_2$[29] also justifies this remark. One may hold the electronic instability in the band nesting states accountable for this peculiarity. We assume that, large proportion of the electron-hole plasma generated at this blue regime (photo-excited with high energy 370 nm pump), would migrate to the lower energy valleys (K, Λ) through inter-valley phonon scattering, and contributes towards the formation of A, B excitons.

Figure 2c depicts the growth profiles of different transient excitonic bleaches. Exciton dynamics are monitored by tuning the probe wavelength at the photoinduced bleach maxima, i.e., band-edge of different excitonic states. We found that there exists a distinct time offset in the growth dynamics of these four excitons, where high energy excitons appear at much faster time scale (at least 0.45 ps) as compared to that of A, B. In addition to that, there is slight time difference in between C and D also. D. Kozawa et al. reported independent inter-valley thermalization phenomena for A exciton emission in bilayer $MoS_2$ and $WS_2$ with excitation of the photo carriers at the band nesting region (C-resonant excitation)[24]. In the present investigation we are exciting the system with much higher energy as compared to that report. Henceforth we expect generation of hot electrons closer to Γ plateau and we cannot neglect the possibility of inter-valley transfer in A/B excitonic growth. So the slow rise time of the A/B excitonic signals would be resultant of both intra-valley (within K-valley itself) and inter-valley thermalization process (from Γ states as well as band nesting region to K space). Very low spectral amplitude of C, D intrigues us to assume that inter-valley thermalization triumphs over the intra one, influencing this extended growth time of A/B. Significant enhancement of A excitonic signal rise time with excitation higher or close to C, D states, confirms immense dependence of A/B exciton formation over Γ-K inter-valley scattering (**Supplementary figure 8a**). For high energy excitons C, D, rise time would refer to mainly intra-valley thermalization where photoexcited carriers form bound pairs, followed by fast relaxation to the excitonic band edge lying in the similar valley or hill as that of photoexcited carriers. As pump excitation energy in the present study is very close to the C, D energy levels, the contribution of the intra-valley scattering process would be minimal in thermalization towards the band edge. Detailed spectral investigations were carried out to further comprehend the formation and relaxation characteristics of these excitons.

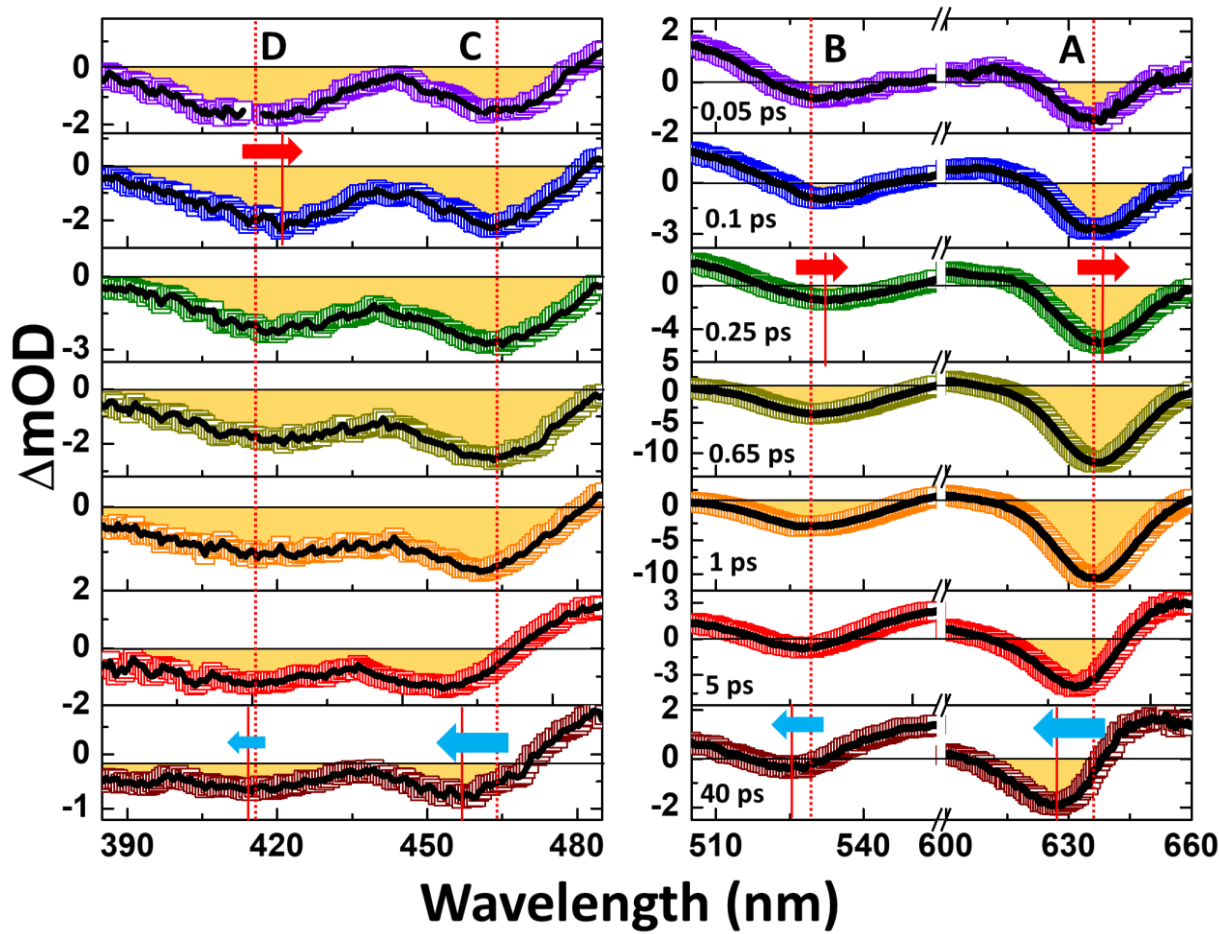

**Figure 3: Transient evolution of excitonic bleach spectra:** Left panel is high energy C, D excitonic bleach spectra and right panel is A, B excitonic bleach spectra following 3.3 eV pump excitations. Temporal evolutions of the excitons are exhibited according to representation of bleach spectra at increasing pump-probe delay time (Downwards: 0.05, 0.1, 0.25, 0.65, 1, 5 and 40 ps). The red dotted lines are starting positions of the respective excitonic peaks, imminently after the photoexcitation. Red and blue arrows indicate corresponding red and blue shift of the spectra.

**Chronology of excitonic spectra and inter-excitonic behaviours.** Figure 3 illustrates temporally and spectrally resolved transient absorption data for all four excitons, non-resonantly pumped at 370 nm. The above scheme can simultaneously monitor excitonic behaviour in both high energy and low energy regime at time delay starting from 0.05 ps to

40 ps. We found that the formation and relaxation behaviours of C, D are very different than those of A, B. Let us first discuss temporal development of A, B. We can divide their spectral variation into three kinetic regimes, I. at 0.05 ps ≤ Δt ≤ 0.25 ps red shift with spectral growth, II. again at 0.25 < Δt ≤ 0.65 ps blue shift with spectral growth, and III. after Δt > 0.65 ps blue shift with spectral decay. So, the growth profiles of these excitons are not a single component, rather it comprises of two distinct processes. Sie et al. proposed similar red and blue shift in rise time of monolayer $WS_2$ and assigned them as hot exciton formation and subsequent thermalization applying nearly identical excitation energy[37]. With the formation of an exciton, optical band gap is decreased owing to intense coulombic interaction within electron and hole (Band gap renormalisation, BGR[38]). Immediately after the formation, hot excitons start cooling towards the K band edge. As excitons start populating the lower energy levels, the apparent optical gap is pushed towards higher levels (also known as Bernstein-Moss effect[39]). Interestingly, both A and B exciton formation and cooling processes occur at very similar time scale, suggests that the hole cooling process within two valence bands occurs almost instantaneously, unlike we observed in case of monolayer $MoS_2$ in our earlier study[40]. Identical energy shifting behaviours of A, B excitons point towards the existence of profound inter-excitonic interactions[41]. Resonant excitation in A, B confirmed this inter-excitonic dependence among each other (**Supplementary figure 7b**).

In high energy regime, excitons exhibit contrasting spectral shifting phenomena amongst themselves. Both excitons reach their respective bleach maxima nearly at ~ 0.25 ps. D exciton is primarily red-shifted followed by a marginal blue-shift during its growth and little blue-shifted while decaying, more like A, B. If we assume that high energy excitons could also behave like A, B, one can interpret that the red shift in growth would be D exciton formation in the hot states and/or coulombic interaction mediated band gap shrinkage, which heads towards thermalization associated blue shift. Very little thermalization time makes sense as D excitonic states would reside very close to the energy states, where the photocarriers are injected. From this analysis we propose that D excitonic states may not lie close to Λ valley or form in a parallel band arrangement, rather find an extreme near Γ hill. Contrary to D, C excitonic bleach exhibit blue shift throughout its temporal evolution, both in its rise and decay process. C exciton happens to be a distinct entity amongst all these excitons that one may think that the photophysical processes behind its shifting behaviours would be different than others. Absence of any red shifting in its growth, points to the instantaneous formation of C excitons. The possible reason behind this blue shifted growth in C would be its immediate dissociation of C excitons via band nesting effect in the parallel band arrangement. With increasing concentration of charge carriers Bernstein-Moss effect or dynamic screening of the coulombic interactions dominates over BGR mediated band gap shrinkage (red shift)[42]. From these observations we can conclude that C, D formation characteristics are independent in nature and unable to influence each other much, unlike A, B, although they are much more closely spaced in energetic terms.

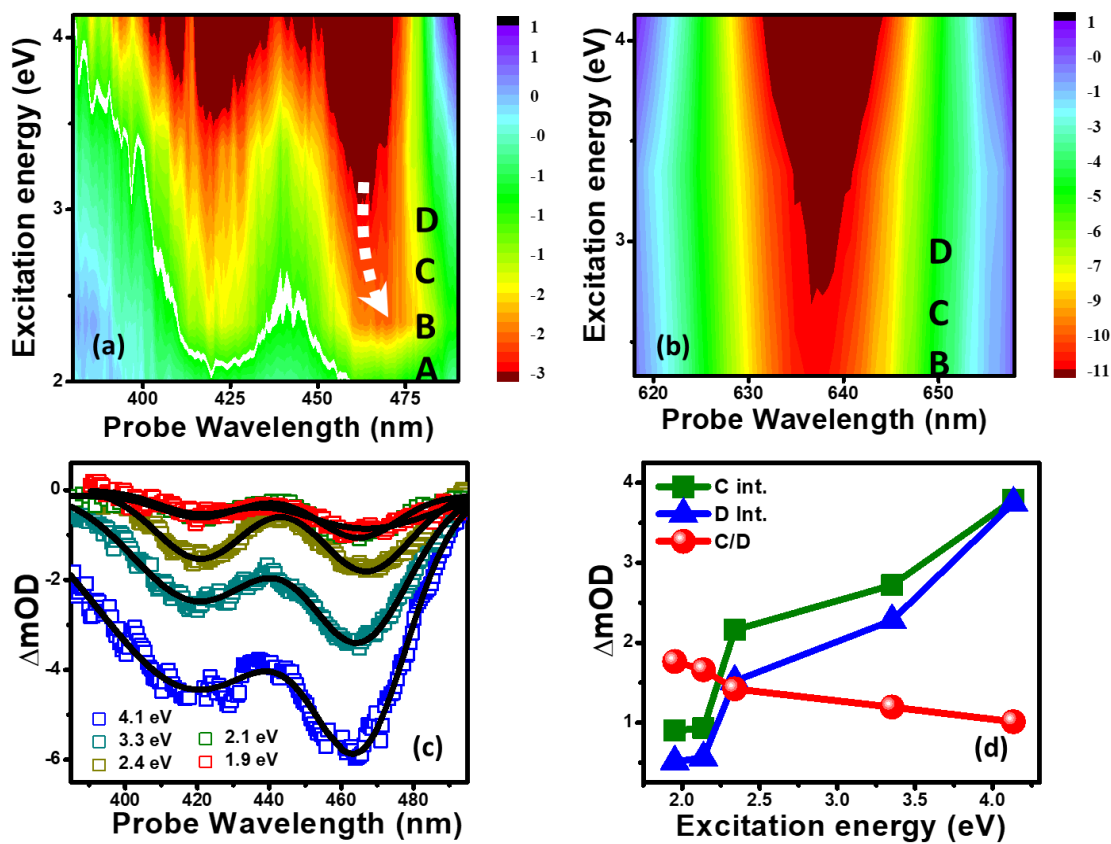

**Figure 4: Transient evolution of high energy excitons with different pump excitation energy.** 2D intensity map of transient absorption spectra at 0.25 ps, simulated from different energy pump excitations for **(a)** C, D and **(b)** A exciton. **(c)** Transient absorption spectra of high energy C, D exciton at 0.25 ps delay time for pump energy 1.9, 2.1, 2.4, 3.3 and 4.1 eV. The solid lines represent Gaussian fitting of the experimental data points. (d) C/D intensity ratio plotted as a function of excitation energy. Individual C, D exciton intensity is also plotted for the reference.

**Exciton formation on pump energy variation.** The intensity map plotted in Figure 4a, b illustrates the evolution of the D, C, A exciton respectively, for a range of pump excitation (1.95-4.13 eV) (B exciton is excluded as it behaves more like A, **Check supplementary figure 7**). The high energy excitons are still visible even on application of much lower excitation energy. Significant overlap in between C excitonic states and K-space excited states makes this up-conversion possible[28,43]. The exciton intensities in all three cases deplete with the decrement of excitation energy. This behaviour correlates well with the fact that, high excitation energy is capable of populating with higher number of charge carriers influencing the phase space filling, responsible for the excitonic bleach formation[38]. This is also true for C excitons where supposed to be no extreme (valley/hill) involved, intrigues us to think that band nesting effect may not be that instantaneous as thought earlier. Interestingly, the C excitonic bleach undergoes a substantial red shift (shown with a dotted arrow in figure 4a) with lowering of pump energy, contrary to other excitons. We attribute this peculiarity to the formation of C exciton in the band nesting region of the brillouin zone. Unlike other excitons, C exciton formation occurs within parallel band arrangement in between Λ-Γ space and is not involved with any extreme points in the conduction and valence band structure. For lower energy pump excitations photoexcited carriers near K-space travel through Λ region to reach C, D excitonic states (**Supplementary Figure 6**). As

C excitons have no particular position (valley minima or hill maxima) in the momentum space, incoming charge carriers fill the lower energy side of the band nesting region in order to construct an electron-hole pair. Charge carriers ejected from the lower energy pump can place itself in the lower side of the parallel band and generate C excitons with a significant red shift. In fact, this energetic transition well-corroborates with GW-Bethe Salpeter calculations for $MoS_2$[43], where the authors discussed C excitonic fine structure formation in diverse regimes of momentum space. D excitons exhibits nearly zero shift with the change of excitation energy, more like A exciton, which intrigues us towards two fundamental conclusions regarding D excitons. One, D exciton does not necessarily form in a parallel band arrangement like C. Two, D exciton formation process sounds more like A, B exciton formation involving extreme points in energy band structure, as expected earlier.

Figure 4c describes evolution of excitonic intensity and broadening of transient absorption spectra of C, D excitons in earlier mentioned range of photoexcitation energy. It is evident that the shape of the transient spectra merely changes with excitation energy. Although, with decrement of excitation energy, C excitonic peaks get broader while D gets narrower (**supplementary figure 6f**). This broadening in case of C excitons reaches maximum at 1.9 eV excitation (lowest energy pump employed in our study). We observed that for high energy excitations (4.1 and 3.3 eV) D excitons are broader than C, whereas as the excitation energy goes down the inverse trend sets in (**Supplementary Figure 6**). We expect this as a resultant of the involvement of lower energy states in the band nesting regime with the formation of C bound pairs in low energy excitations. The intensity of C and D excitons are extracted and plotted as a function of pump excitation energy in figure 4d. We used maximum bleach intensity as it directly correlates maximum exciton population in the respective excitonic band edge. As we decrease the pump photon energy, intensity of the bleach signals decreases; D exciton intensity decreases at much faster rate than C. An almost linear increment of C/D bleach intensity ratio was observed in our pump photon energy range. This leads us to the fact that, with lower energy pump C exciton formation probability is much higher than D. The reason behind this could be the incapability of lower energy pump in providing hot carriers into the D excitonic states.

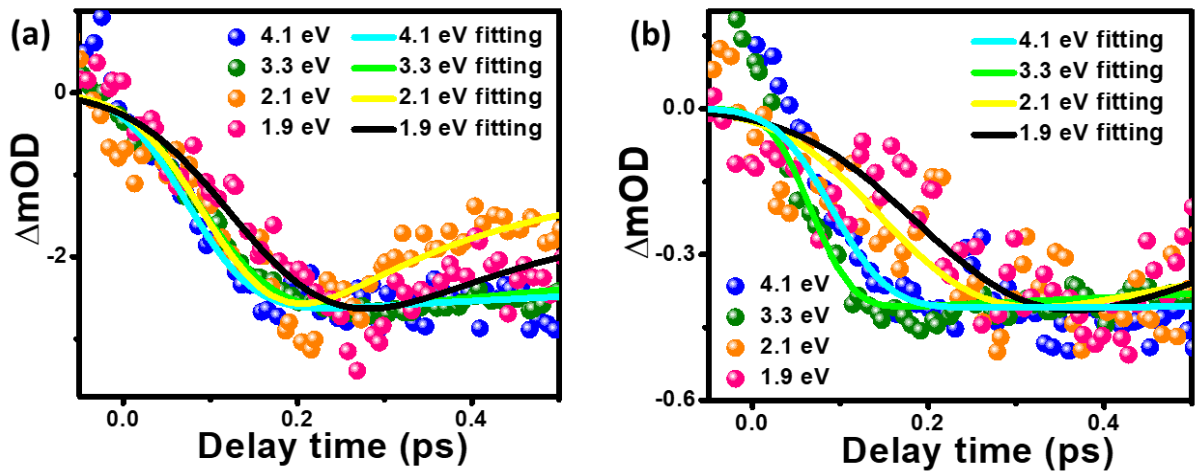

**Figure 5: Transient kinetics of the high energy excitons with different excitation energy.** Transient Growth profiles of **(a)** C and **(b)** D excitons with application of 4.1, 3.3, 2.1 and 1.9 eV pump energy, with similar fluence used earlier.

**High energy excitonic maneuvers on variation of pump energy.** To analyse different pump energy induced effect on the formation of $WS_2$ excitons, spatially and energitically dispersed over a wide range, we tried to compare their formation characteristics for few specific energy excitations. Figure 5a, b presents growth features of C, D excitons respectively, with 4.1, 3.3, 2.1 and 1.9 eV pump energy excitation, which will be further reffered as higher energy excitation, HE (4.1 and 3.3 eV) and lower energy excitation, LE (2.1 and 1.9 eV). C, D bleach maxima experience anomolous change in rise time much unlike A (**Supplementary Figure 8**) with increasing excitation energy. For all the excitations, C exhibits almost identical rise time, while there is a distinct difference the rise times for D at different excitation energy. In order to explain these events we need to understand how HE and LE excitations work for C, D formation. In case of HE, photocarriers are excited in the high energy levels, from where they thermalize themselves at the excitonic band edges. Whereas in LE scenario, C, D states are populated from K excited state mediated tunneling. So, the growth time observed here would be collective potrayal of excitation of carriers in the K space and subsequent tunneling. As there should not be much diffrence in tunneling time, probe mediated absorption of photocarriers into the C-hybridised K states from the lower energy states in K-valley would be responsible for the observed growth process of C excitationic signal in LE. For HE excitation C rise time is a compilation of immediate formation of C excitons and opposite propagation of electron and hole in the parallel band arrangement, as dicussed earlier. Identical rise time for 3.3 and 4.1 eV excitation implies that band nesting propagation does not necessarily depend on the excitation energy. Whereas, on increasing photoexcitation energy from 3.3 to 4.1 eV, D excitons introduce enhancemement in the signal rise time, in contrast to the indifferent behaviour of C. This indicates that, D exciton formation is not instanteneous and affected with variation of excitation energy more like A, B. In LE case, D exciton growth should be affected more as it resides in higher energitical position. That's why we observed much increase in D exciton rise time for 2.1 and 1.9 eV photo-excitation. In fact this large growth time scale in D (even higher than any excitation of C) proposes that, there is no excitonic overlap within D states and K high energy states and K-Γ tunneling to D states is not viable.

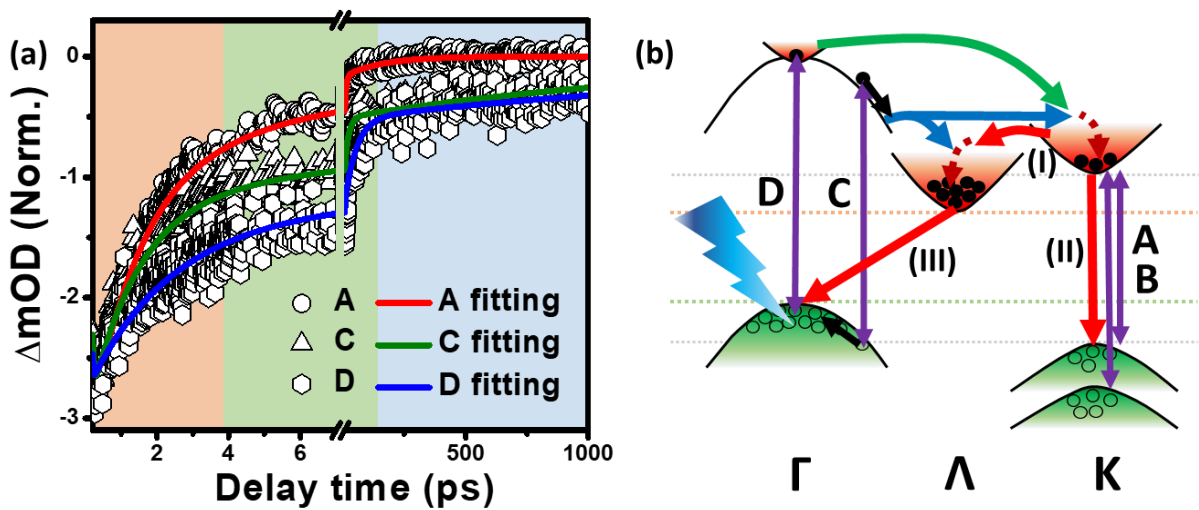

**Figure 6: Transient decay of excitonic bleach features. (a)** Normalized kinetic traces of A, B, C, D probed at excitonic bleach maxima (636, 532, 464 and 422 nm) following 3.3 eV photoexcitation. The solid lines are multiexponential fitting results of the experimental data. **(b)** Schematic optical valley diagram of few layer WS$_2$ and possible relaxation pathways for the excitons following 370 nm photoexcitation. Here, only Γ space excitations are considered for the sake of simplicity. Also, less-probable processes are omitted here. Green arrows represent inter-valley electron transfer from D states, whereas blue arrows symbolize electronic delocalisation of C excitons into Λ and K valley. C excitonic propagation due to band nesting effect is shown with two antiparallel black arrows for electron and hole. Red arrows present subsequent migration of electrons (common for all the excitons), where (I) inter-valley scattering of electrons from K to Λ valley, (II) K-K direct recombination and (III) Λ-Γ indirect recombination. Red dotted lines are intra-valley thermalization processes in K and Λ valley.

**Relaxation pathway in few layer WS$_2$.** In figure 6a transient kinetic profile of exciton A is directly compared with that of high energy excitonic features (C, D). These decay profiles were fitted with multi-exponential time constants and are shown in supplementary table 1 where all the fitting parameters are listed. The decay time components $\tau_2$, $\tau_3$, and $\tau_4$ are also marked in three different background colours in figure 6a. B exciton dynamics is excluded from the discussion, as it is strongly influenced by the surrounding photo-induced absorption signals (**supplementary figure 9**). The decay dynamics of C and D excitonic signals, are much slower as compared to A, despite having very low spectral weight. Previously few articles reported similar slow relaxation of C excitons compared to A/B in monolayer MoS$_2$, and elucidated that stifled inter-valley electron transfer from Λ to K valley would be responsible for the retarded nature of C transient signal[28,29]. However, in few layer case the scenario would not be the same. Here, the electronic band structure modifies itself with increasing layers, placing CBM at Λ valley and VBM in Γ hill, in contrast to the monolayer TMDC. This adversely affect the direct K-K transition, favouring indirect Λ-Γ and K-Γ transitions. Now, in case of C excitons, following the formation of the bound pair, the electron and hole start propagating in opposite direction towards Λ and Γ respectively (Band nesting effect). This migration of electrons could end up in Λ valley, owing to closer proximity of C excitonic states with Λ valley compared to K and subsequently being thermalized to the Λ valley edge. However, C electrons could also easily migrate towards the K regime as C excitonic states and K excited states enjoy significant overlap[28,43], as we have shown in our previous discussion. Hot electrons in K valley would either be transferred into the Λ valley through K-Λ inter-valley scattering or would undergo intra-valley thermalization in K valley itself and help in formation of A/B excitons. In room temperature photoluminescence (PL) measurements for different layers of $WS_2$, Zhao et al. showed that indirect Λ-Γ PL peaks are enhanced with increasing layer nos, however the PL sepctrum is also comprised of PL peaks corresponding to the direct K-K recombination[32]. However, they failed to observe any indirect PL peak corresponding to the K-Γ recombination at room temperature. From this instance, we can assume that Λ-Γ indirect pathway may be the most feasible channel for the relaxation of $WS_2$ excitons, although K-K direct recombination cannot be neglected. D excitons would also end up in K valley rather in Λ, as Γ-K inter-valley scattering would be much stronger than the Γ-Λ one, as K valley lies closer to Γ valley than that of Λ[32]. Subsequently D excitons would follow C relaxation pathway. In figure 6b we presented a simple energy valley schematics of few layer $WS_2$ portraying electronic transitions corresponding to all four excitonic features following 370 nm photo-excitation (**See supplementary figure 9 discussion for more details**).

From multi-exponential fitting, decay time components for the transient bleach signals were found to be, $\tau_2$ ~ 1.4 ps (62.2 %), $\tau_3$ ~ 15.8 ps (18.6 %) and $\tau_4$ > 1 ns (19.2 %) for C excitons and $\tau_2 \sim 2.4$ ps (53.3 %), $\tau_3 \sim 46.6$ ps (27.8 %) and $\tau_4 > 1$ ns (18.9 %) in case of D. The slow decay component ($\tau_4$) possess a large time scale value (> 1 ns) and resembles with retarded nature of indirect recombination processes observed in bulk samples[28,44]. Hence, we can consider the $\Lambda$-$\Gamma$ indirect recombination pathway to be the primary channel for C, D relaxation. Based on our previous discussion we ascribe $\tau_2$, $\tau_3$ and $\tau_4$ as inter-valley electron transfer ($\Gamma$-K-$\Lambda$ for D and K-$\Lambda$ for C), intra-valley thermalization in $\Lambda$ and $\Lambda$-$\Gamma$ indirect recombination, respectively. $\tau_2$ and $\tau_3$ is much slower in case of D excitons than that of C, mainly because of pauli blocking effect in K valley in course of $\Gamma$(D)-K inter-valley electron transfer and in intra-valley thermalization process, as the states are already filled from either C or A/B electrons. Contrary to C and D, A exciton exhibit very fast decay ($\tau_4 \sim 150$ ps). This intrigues us to think that, K-K direct recombination process is the significant contributor for the relaxation of A[44]. The possible reason would be the lower probability of K-$\Lambda$ inter-valley scattering for K band edge electrons, owing to less kinetic energy of band edge electrons as compared to the high energy ones[45]. Although we could not analyse B exciton dynamics separately, we expect relaxation of B would follow A excitons, as these two are immensely dependent on each other.

**Discussion**

In atomically thin TMDC systems, high energy excitons are capable of absorbing impressive amount of electromagnetic radiation contrary to other layered materials having similar thickness, riding on their high oscillator strength. Proper utilization of these properties may open new avenues in the 2D opto-electronics market. Basic understanding of the formation and relaxation behaviours of high energy excitons (C, D) in a TMDC system for varient optical excitations is very important, in order to use those materials in broadband effective opto-electronic application. Optical behaviour of a few layer TMDC system falls in between direct monolayer and indirect bulk entitities. In few layer case almost all the optical signatures of monolayers are retained despite Λ-Γ being the optical indirect gap and act as an key recombination channel more like bulk. We observed instanteneous formation of high energy excitonic species, followed by slugghish relaxation as compared to their K-space counterparts even after lower energy (LE) excitation. K-excited state mediated ultrafast tunneling into C excitonic states works as the primary route for the formation of C in LE. However, D states do not possess any excitonic overlap with K excited states and mainly populated from photo-induced absorption of carriers in C states.

We elucidated that, D excitons emerge close to Γ region, finds a valley-hill arrangement in the energy band, contrary to C exciton formation in a parallel band. C, D exciton depletion could inject large amount of electrons in the K valley, contributing in A, B exciton population in K space. Moreover, formation of A, B excitons also influence in populating C, D states irrespective of the excitation energy. However, relaxation characteristics of C, D excitons are completely different from that of A, B. C, D relaxation follows the bulk route, first inter-valley electron transfer from Γ to k valley and K to Λ valley, subsequently slower intra-valley thermalization in Λ valley and Λ-Γ indirect recombination. Involvement of inter-valley scattering processes retard the relaxation of these high energy excitons. Pauli blocking in intra-valley thermalization and slow indirect recombination channel also serves as a bottleneck in C, D relaxation pathway and extends their charge separation. D excitons are most severly affected from the pauli blocking in both K and Λ valley and results in very slow relaxation. Contrary to these high energy excitons, the primary relaxation channel for A, B was found to be direct K-K recombination. Trap mediated non-radiative recombination in K space help in fast relaxation of A, B excitons. Large absorption coffecient and slow realaxation behaviour of high energy excitons would have immense effect in fabricating few layer TMDC based broadband opto-electronic devices, where exctraction of hot carriers play vital role.

In summary we explained the formation and corresponding relaxation behaviours of high energy excitonic features, in mirror of their low energy counterparts for few layer $WS_2$. In course of that, we prepared $WS_2$ nanosheets with 5-6 no of layers and observed four independent excitonic peaks in steady state optical absorption spectrum. Our ultrafast measurements reveal weak C, D excitonic bleach formation at the blue region of the transient spectra along with fundamental A, B bleach signal at the red region irrespective of the applied excitation energy. Large exciton binding energy due to greater electron hole effective mass in the Λ, Γ region[25,32] ensures their existance even in the weak confined system like few layer. With high energy excitation, large no of photo-excited electrons generated in the Γ space undergo Γ-K inter-valley transfer even before the formation of C, D bound pairs and greatly influence A, B exciton formation. D exciton emerges in a valley-hill arrangement close to Γ, more like A, B and shows no effect of band nesting, unlike C, which forms bound pair in the band nesting region in between Λ-Γ. We elucidated that, C exciton does not possess any particular momentum space position in the optical band diagram, betting on pump excitation energy C exciton may emerge in a wide range of momentum space within Λ-Γ, establishing previous theoritical predictions. C excitons immediately dissociate in the parallel bands via a opposite and equal propagation of electron and holes towards Λ valley and Γ hill through band nesting effect. C, D excitons relax through mainly Λ-Γ indirect pathway, unlike fast K-K recombination in A, B. We expect our detailed analysis on excitonic features in a broad range of the electromagnetic spectrum for a few layer system would inspire more studies on valley based exciton formations in 2D materials and help in constructing unique 2D opto-elctronic devices.

# Methods

**Sample preparation.** Bulk $WS_2$ powder and N-Methyl-2-Pyrrolidone (NMP) were purchased from Sigma-Aldrich. Liquid exfoliation technique was employed to prepare few layer $WS_2$ nanosheets. First, 80 mg $WS_2$ powder was crushed with the help of mortar and pestle for 30 mins to make fine uniform powder. These crushed powders were solvated in 15 ml of NMP and subsequently sonicated for 4h in an ultrasonic bath sonicator. Then, dispersions were probe sonicated for 1h and centrifuged at 4000 rpm for 30 mins. The above three-fourth part of supernatant containing few layers of $WS_2$, were carefully pipetted off and collected for further studies.

**Femtosecond transient absorption spectroscopy.** The ultrafast setup was comprised of a Ti: sapphire amplifier system (Astrella, Coherent, 800 nm, 3mJ/pulse energy, ~ 35 fs pulse width and 1 kHz repetition rate) and Helios Fire pump-probe spectrometer. With the help of two beam splitters the output laser pulses were cleaved into pump (95% of the output) and probe (remaining 5%) beams. Optical Parametric Amplifier (OPerA-SOLO) was in place of pump beam to produce required wavelength for the photo-excitation of the sample. A delay stage is operating in path of probe beam to maintain a perfect delay between pump and probe in temporal sense throughout the experimental procedure. Monochromatic probe light passes through a $CaF_2$ crystal generating UV-Visible probe pulses. Probe beam passes through sample dispersion and falls upon fibre coupled CMOS detectors connected with the computer system. All the experimental measurements were carried out at room temperature. The collected TA data were chirped and fitted multi-exponentially using surface explorer software.

**Acknowledgements**

T.G., H. B., K. J. B, G. K. and N. G. are grateful towards Institute of Nano Science and Technology (INST), Mohali, India for the research fellowship. The authors acknowledge INST, Mohali, India for providing instrumental facility and supporting this research work.

**Author contribution**

T. G. performed the experiments and analysed the experimental data. H. B. and K. J. B. helped in synthesizing and characterising the material. G. K. and N. G. performed the detailed spectral analysis and helped with the transient experiments. T.G. and H. N. G. wrote the paper.

**Competing interests**